\documentclass[12pt]{article}
\usepackage{a4wide}

\newcommand{\DD}{{\cal D}}
\newcommand{\EE}{\hbox{\it I\kern-3pt E}}
\newcommand{\LL}{{\cal L}}
\newcommand{\OO}{{\cal O}}
\newcommand{\PP}{{\cal P}}
\newcommand{\TT}{{\cal T}}
\newcommand{\RR}{\hbox{\it I\kern-3pt R}}
\newcommand{\VV}{{\cal V}}

\newcommand{\Ad}{\mathop{\rm Ad}\nolimits}
\newcommand{\dirvec}{\rlap{\kern-3pt\raise2pt\hbox{$^{^{\ \rightarrow}}$}}}
\newcommand{\revvec}{\rlap{\kern-3pt\raise2pt\hbox{$^{^{\ \leftarrow}}$}}}
\newcommand{\dubvec}{\rlap{\kern-3pt\raise2pt\hbox{$^{^{\ \leftrightarrow}}$}}}
\newcommand{\slD}{\kern1pt/\kern-8pt D}
\newcommand{\slF}{F\kern-8pt/\kern2pt}
\newcommand{\slG}{G\kern-8pt/\kern2pt}
\newcommand{\slk}{/\kern-6pt k}
\newcommand{\oone}{\hbox{\rm 1\kern-3pt l}}

\begin{document}

\begin{flushright}
MZ-TH/10-20\\
June 2010
\end{flushright}

\begin{center}
{\Large\bf Gauge and Lorentz transformation\\[7pt]
  placed on the same foundation}

\vspace{24pt}
{\large R.~Saar$^1$, S.~Groote$^{1,2}$, H.~Liivat$^1$ and I.~Ots$^1$}\\[12pt]
$^1$ Loodus- ja Tehnoloogiateaduskond, F\"u\"usika Instituut,\\
  Tartu \"Ulikool, Riia~142, 51014 Tartu, Estonia\\[12pt]
$^2$ Institut f\"ur Physik der Johannes-Gutenberg-Universit\"at,\\
  Staudinger Weg 7, 55099 Mainz, Germany
\vspace{24pt}
\end{center}


\begin{abstract}
In this note we show that a ``dynamical'' interaction for arbitrary spin can
be constructed in a straightforward way if gauge and Lorentz transformations
are placed on the same foundation. As Lorentz transformations act on
space-time coordinates, gauge transformations are applied to the gauge field.
Placing these two transformations on the same ground means that all
quantized field like spin-$1/2$ and spin-$3/2$ spinors are functions not only
of the coordinates but also of the gauge field components. This change of
perspective solves a couple of problems occuring for higher spin fields like
the loss of causality, bad high-energy properties and the deviation of the
gyromagnetic ratio from its constant value $g=2$ for any spin, as caused by
applying the minimal coupling. Starting with a ``dynamical'' interaction, a
non-minimal coupling can be derived which is consistent with causality, the
expectation for the gyromagnetic ratio, and well-behaved for high energies. 
As a consequence, on this stage the (elektromagnetic) gauge field has to be
considered as classical field. Therefore, standard quantum field theory cannot
be applied. Despite this inconvenience, such a common ground is consistent
with an old dream of physicists almost a century ago. Our approach, therefore,
indicates a straightforward way to realize this dream.
\end{abstract}
\begin{center}
{\em published as Advances in Mathematical Physics 2011 (2011) 652126}
\end{center}

\newpage

\section{Introduction}
After the formulation of general relativity which explained fources on a
geometric ground, physicists and mathematicians tried to incorporate the
electromagnetic interaction into this geometric picture. Hermann Weyl claimed
that the action integral of general relativity is invariant not only under
space-time Lorentz transformations but also under the gauge transformation, if
this is incorporated consistently~\cite{Weyl:1919}. However, the theories at
that time were not ready to incorporate this view. Nowadays, we see more
clearly that all physical variables (like position, momentum, etc.), quantum
wave functions and fields transform as finite-dimensional representations of
the Lorentz group. The reason is that interactions between fundamental
particles (as irreducible representations of the Poincar\'e group) are most
conveniently formulated in terms of field operators (i.e., finite-dimensional
representations of the Lorentz group) if the general requirements like
covariance, causality, etc.\ are to be incorporated in a consistent way. The
relation between these two groups and their representations is given by the
Lorentz-Poincar\'e connection~\cite{Tung:1985na}. In this note we show that if
gauge transformation is put on the same foundation, the resulting non-minimal
``dynamical'' interaction obeys all necessary symmetries which for higher
spins are broken if the interaction is introduced by the usual minimal
coupling.

\vspace{7pt}
In Sec.~2 we explain details of the Poincar\'e group which are necessary in
the following. In Sec.~3 we deal with linear wave equations as objects to the
Lorentz transformation. In Sec.~4 we introduce the external electromagnetic
field by a nonsingular transformation. In Sec.~5 we specify the nonlinear
transformation by the claim of gauge invariance of the Poincar\'e algebra.
Finally, in Sec.~6 we give our conclusions.

\section{The Poincar\'e group}
Relativistic field theories are based on the invariance under the Poincar\'e
group $\PP_{1,3}$ (known also as inhomogeneous Lorentz group
${\cal IL}$~\cite{Wigner:1939cj,Bargmann:1954gh,Fronsdal:1959zz,Shaw:1964zz,%
Joos:1962qq,Niederer:1974ps,Ohnuki:1988ai,Tung:1985na,Kim:1986wv}) This group
is obtained by combining Lorentz transformations $\Lambda$ and space-time
translations $a_T$,
\begin{equation}
(a,\Lambda)\equiv a_T\Lambda:\EE_{1,3}\ni
  x^\mu\to{\Lambda^\mu}_\nu x^\nu+a^\mu\in\EE_{1,3}.
\end{equation}
The group's composition law
$(a_1,\Lambda_1)(a_2,\Lambda_2)=(a_1+\Lambda_1a_2,\Lambda_1\Lambda_2)$
generates the semidirect structure of $\PP_{1,3}$,
\[\PP_{1,3}=\TT_{1,3}\odot\LL\]
where $\TT_{1,3}$ is the abelian group of space-time translations (i.e.\ the
additive group $\RR^4$) and $\LL=\{\Lambda:\det\Lambda=+1,{\Lambda^0}_0\ge 1\}$
is the proper orthochronous Lorentz group acting on the Minkowski space
$\EE_{1,3}$ with metric
\[\eta_{\mu\nu}=\mathop{\rm diag}(1,-1,-1,-1).\]
The condition of the metric to be invariant under Lorentz transformations
$\Lambda$ takes the form
\begin{equation}\label{eq02}
{\Lambda^\mu}_\rho\eta_{\mu\nu}{\Lambda^\nu}_\sigma=\eta_{\rho\sigma}.
\end{equation}
Under the Lorentz transformation $\Lambda\in\LL$ the transformation of the
covariant functions $\psi$ according to a representation $\tau(\Lambda)$ of
the Lorentz group~\cite{Wigner:1939cj,Bargmann:1948ck,Corson:1982wi,%
Pursey:1965zz,Tung:1967zz,Wightman:1977uc,Bargmann:1954gh,Fronsdal:1959zz,%
Shaw:1964zz,Joos:1962qq,Niederer:1974ps} is determined by the commutative
diagramm
\begin{center}\begin{tabular}{rlcr}
$\psi:$&$x\in\EE_{1,3}$&$\longrightarrow$&$\psi(x)$\\
$\llap{$\tau(\Lambda)$}\downarrow$&$\downarrow\rlap{$\Lambda$}$&&
  $\downarrow\rlap{$T(\Lambda)$}\quad$\\
$\tau(\Lambda)\psi:$&$\Lambda x$&$\longrightarrow$&$T(\Lambda)\psi(x)$\\
\end{tabular}\end{center}
i.e.\
\begin{equation}\label{eq01}
T(\Lambda)\psi(x)=(\tau(\Lambda)\psi)(\Lambda x)\equiv\psi^\Lambda(\Lambda x).
\end{equation}
The map $T:\Lambda\to T(\Lambda)$ is a finite-dimensional representation of
$\LL$. If we parametrize the element $\Lambda\in\LL$ by
$\Lambda(\omega)=\exp(-\frac12\omega_{\mu\nu}e^{\mu\nu})$ where the Lorentz
generators are given by
\[
{(e_{\mu\nu})^\rho}_\sigma=-{\eta_\mu}^\rho\eta_{\nu\sigma}+\eta_{\mu\sigma}
{\eta_\nu}^\rho
\]
and $\omega^{\mu\nu}=-\omega^{\nu\mu}$ are six independent parameters, the
parametrization of $T$ reads
\[
T(\Lambda(\omega))=\exp\Big(-\frac i2\omega_{\mu\nu}s^{\mu\nu}\Big).
\]
The Lorentz group $\LL$ is non-compact. As a consequence, all unitary
representations are infinite dimensional. In order to avoid this, we introduce
the concept of $H$-unitarity (see e.g.\ Ref.~\cite{Niederer:1974ps} and
references therein). A finite representation $T$ is called $H$-unitary if
there exists a nonsingular hermitian matrix $H=H^\dagger$ so that
\begin{equation}\label{eq03}
T^\dagger(\Lambda)H=HT^{-1}(\Lambda)\quad\Leftrightarrow\quad
s_{\mu\nu}^\dagger H=Hs_{\mu\nu}.
\end{equation}
Notice that a $H$-unitary metric is always indefinite, so that the inner
product $\langle\ ,\ \rangle$ generated by $H$ is sesquilinear sharing the
hermiticity condition
$\langle\psi,\varphi\rangle=\langle\varphi,\psi\rangle^*$. The most famous
case of $H$-unitarity is given in the Dirac theory of spin-1/2 particles
where $H=\gamma^0$.

\vspace{7pt}
For an operator $\OO$~\cite{Giovannini:1977wi,Janner:1970zz} acting on the
$\psi$-space of covariant functions\footnote{We have to impose the action on
covariant functions because in case of higher spins the relations between
operators we obtain are valid only as weak conditions.} the transformation
$\tau(\Lambda)$ in Eq.~(\ref{eq01}) is a covariant transformation if the
diagramm
\begin{center}\begin{tabular}{rlcr}
$\OO\psi:$&$x$&$\longrightarrow$&$(\OO\psi)(x)$\\
$\llap{$\tau(\Lambda)$}\downarrow$&$\downarrow\rlap{$\Lambda$}$&
  &$\downarrow\ T(\Lambda)$\\
$\tau(\Lambda)(\OO\psi):$&$\Lambda x$&$\longrightarrow$
  &$T(\Lambda)(\OO\psi)(x)$\\
\end{tabular}\end{center}
is commutative, i.e.\
\begin{equation}\label{eq05}
(\tau(\Lambda)\OO\tau^{-1}(\Lambda))(\Lambda x)(\tau(\Lambda)\psi)(\Lambda x)
  =T(\Lambda)\OO(x)\psi(x).
\end{equation}
Using Eq.~(\ref{eq01}) we obtain
\[(\tau(\Lambda)\OO\tau^{-1}(\Lambda))(\Lambda x)T(\Lambda)\psi(x)
  =T(\Lambda)\OO(x)\psi(x).\]
Notice that the covariance of the transformation embodies only the property of
equivalence of reference systems. The covariant operator $\OO$ is invariant
under the transformation~(\ref{eq01}) if in addition
$\tau(\Lambda)\OO\tau^{-1}(\Lambda)=\OO$.
As a consequence we obtain the commutative diagram
\begin{equation}\label{kom0}
\begin{tabular}{rlcr}
$\OO\psi:$&$x$&$\longrightarrow$&$(\OO\psi)(x)$\\
$\llap{$\tau(\Lambda)$}\downarrow$&$\downarrow\rlap{$\Lambda$}$&
  &$\downarrow\ T(\Lambda)$\\
$\OO(\tau(\Lambda)\psi):$&$\Lambda x$&$\longrightarrow$
  &$T(\Lambda)(\OO\psi)(x)$\\
\end{tabular}
\end{equation}
or $\OO(\Lambda x)T(\Lambda)\psi(x)=T(\Lambda)\OO(x)\psi(x)$ which means
\begin{equation}\label{eq06}
\OO(\Lambda x)T(\Lambda)=T(\Lambda)\OO(x)
\end{equation}
on the $\psi$-space. The invariance is a symmetry of the physical system and
implies the conservation of currents. In particular, the symmetry
transformations leave the equations of motion form-invariant.

\vspace{7pt}
While the Lorentz transformation $T(\Lambda)$ changes the wave function $\psi$
itself as well as the argument of this function (cf.\ Eq.~(\ref{eq01})), the
proper Lorentz transformation $\tau(\Lambda)$ causes a change of the wave
function only. On the ground of infinitesimal transformations, this change is
performed by the substancial variation. Starting from an arbitrary
infinitesimal coordinate transformation
$\Lambda(\delta\omega):x^\mu\to x^\mu+\delta\omega^{\mu\nu}x_\nu$, the
substancial variation is given by Ref.~\cite{Corson:1982wi}
\[
\delta_0\psi(x)\equiv\psi'(x)-\psi(x)
  =-\frac{i}2\delta\omega^{\rho\sigma}M_{\rho\sigma}\psi(x)
\]
where $M_{\rho\sigma}=\ell_{\rho\sigma}+s_{\rho\sigma}$,
$\ell_{\rho\sigma}=i(x_\rho\partial_\sigma-x_\sigma\partial_\rho)$. The
corresponding finite proper Lorentz transformation can be written as
\[\tau(\Lambda(\omega))=\exp\left(-\frac{i}2\omega_{\mu\nu}M^{\mu\nu}\right),\]
and the multiplicative structure of the group generates the adjoint action
\begin{equation}\label{adjact}
\mbox{Ad}_{\tau(\Lambda)}:M_{\mu\nu}\to\tau^{-1}(\Lambda)M_{\mu\nu}
  \tau(\Lambda)={\Lambda_\mu}^\rho{\Lambda_\nu}^\sigma M_{\rho\sigma}.
\end{equation}
Due to Eq.~(\ref{eq03}) the generators $s_{\rho\sigma}$ fulfill
$s_{\rho\sigma}^\dagger H=Hs_{\rho\sigma}$. They depend on the spin of the
field but not on the coordinates $x_\mu$. Therefore, we have
$[\ell_{\mu\nu},s_{\rho\sigma}]=0$. If a generic element of the translation
group is written as
\[\exp(+ia_\mu P^\mu),\]
the commutator relations of the Lie algebra are given by
\begin{eqnarray}\label{eq08}
[M_{\mu\nu},M_{\rho\sigma}]&=&i(\eta_{\mu\sigma}M_{\nu\rho}
  +\eta_{\nu\rho}M_{\mu\sigma}-\eta_{\mu\rho}M_{\nu\sigma}
  -\eta_{\nu\sigma}M_{\mu\rho}),\nonumber\\[7pt]
[M_{\mu\nu},P_\rho]&=&i(\eta_{\nu\rho}P_\mu-\eta_{\mu\rho}P_\nu),
  \nonumber\\[7pt]
[P_\mu,P_\nu]&=&0.
\end{eqnarray}
The Casimir operators of the algebra are $P^2=P_\mu P^\mu$ and
$W^2=W_\mu W^\mu$ where
\[W^\mu=+\frac12\epsilon^{\mu\nu\rho\sigma}M_{\nu\rho}P_\sigma\]
is the Pauli-Lubanski pseudovector, $[P_\mu,W_\nu]=0$. In coordinate
representation we have $P_\mu=i\partial_\mu$, and the finite Poincar\'e
transformation has the form
\begin{equation}
\tau(a,\Lambda):\psi(x)\to\left(\tau(a,\Lambda)\psi\right)(x)
  =T(\Lambda)\psi\left(\Lambda^{-1}(x-a)\right).
\end{equation}
This relation constitutes the Lorentz--Poincar\'e
connection~\cite{Tung:1985na}. While the representation $T$ generally
generates a reducible representation of $\PP_{1,3}$, the spectra of the
Casimir operators $P^2$ and $W^2$ determine the mass and spin content of the
system.

\section{The wave equations}
As an operator $\OO$ in the above sense we consider the operator of the
wave equation. The Dirac-type wave equation we will consider has the form
\begin{equation}\label{eq10}
\DD(\partial)\psi(x)\equiv(i\beta^\mu\partial_\mu-\rho)\psi(x)=0
\end{equation}
where $\psi$ is an $N$-component function, $\beta^\mu$ ($\mu=0,1,2,3$), and
$\rho$ are $N\times N$ matrices independent of $x$. Following Bhabha's
conception~\cite{Bhabha:1945zz}, it is ``\dots\ logical to assume that the
fundamental equations of the elementary particles must be first-order
equations of the form~(\ref{eq10}) and that all properties of the particles
must be derivable from these without the use of any further subsidiary
conditions.''

\vspace{7pt}
The principle of relativity states that a change of the reference frame cannot
have implications for the motion of the system. This means that
Eq.~(\ref{eq10}) is invariant under Lorentz transformations. Equivalently, the
Lorentz symmetry of the system means the covariance and form-invariance of
Eq.~(\ref{eq10}) under the transformation in Eq.~(\ref{eq01}), i.e.\ the
transformed wave equation is equivalent to the old one. Therefore, we require
that every solution $\psi^\Lambda(\Lambda x)$ of the transformed equation
\[\DD^\Lambda(\Lambda\partial)\psi^\Lambda(\Lambda x)=0\]
can be obtained as Lorentz transformation of the solution $\psi(x)$ of
Eq.~(\ref{eq10}) in the original system and that the solutions in the original
and transformed systems are in one-to-one correspondence. The explicit form of
the covariance follows from Eq.~(\ref{eq05}),
\begin{equation}\label{eq11}
\left(\tau(\Lambda)\DD\tau^{-1}(\Lambda)\right)(\Lambda\partial)
\left(\tau(\Lambda)\psi\right)(\Lambda x)=T(\Lambda)\DD(\partial)\psi(x)=0
\end{equation}
and leads to the explicit Lorentz transformations
\[\beta^{\Lambda\mu}={\Lambda^\mu}_\rho T(\Lambda)\beta^\rho T^{-1}(\Lambda),
  \qquad\rho^\Lambda=T(\Lambda)\rho T^{-1}(\Lambda).\]
The Lorentz invariance is given by the substitution
\[\DD(\partial)\psi(x)=0\
\buildrel{{\rm Eq.\,(\ref{eq01})}}\over\longrightarrow\
\DD(\partial)\psi^\Lambda(x)=0.\]
or
\[T^{-1}(\Lambda)\beta^\mu T(\Lambda)={\Lambda^\mu}_\rho\beta^\rho,\qquad
T^{-1}(\Lambda)\rho T(\Lambda)=\rho.\]
The difference of the original and transformed wave equation is given by the
wave equation where the wave function $\psi$ is replaced by the substancial
variation $\delta_0\psi$, $\DD(\partial)\delta_0\psi(x)=0$. As a consequence
we obtain $[\DD,M^{\rho\sigma}]=0$ or
\begin{equation}\label{eq13}
[\beta^\mu,s^{\rho\sigma}]=i(\eta^{\mu\rho}\beta^\sigma
  -\eta^{\mu\sigma}\beta^\rho),\qquad[\rho,s^{\rho\sigma}]=0.
\end{equation}
An excellent discussion of such matrices $\beta$ can be found in
Refs.~\cite{Fierz:1939zz,Bhabha:1945zz,Wild:1947zz,Corson:1982wi,Gelfand:1963,%
Naimark:1964}. The hermiticity of the representation $T$ in Eq.~(\ref{eq03})
implies the hermiticity of Eq.~(\ref{eq10}). Including a still unspecified
hermitian matrix $H$ the hermiticity condition reads
$\DD(\partial)^\dagger H\buildrel!\over=(\DD(\partial)H)^\dagger
=H\DD(-\partial)$ or
\begin{equation}\label{eq14}
\beta^{\mu\dagger}H=H\beta^\mu,\qquad\rho H=H\rho.
\end{equation}
Writing $\bar\psi=\psi^\dagger H$, one obains the adjoint equation
\begin{equation}\label{eq15}
\bar\psi\DD(-\revvec\partial)=\bar\psi(-i\beta^\mu\revvec{\partial_\mu}-\rho)
  =(H\DD(\partial)\psi)^\dagger=0.
\end{equation}

\section{Introduction of the external field}
It may be reasonable to introduce an external field directly into the
Poincar\'e algebra which can be applied to classically understand the
elementary particle. To do so one has to transform the generators of the
Poincar\'e group to be dependent on the external field in such a way that the
new, field-dependent generators obey the commutation relations~(\ref{eq08}).
As it was proposed by Chakrabarti~\cite{Chakrabarti:1968zz} and Beers and
Nickle~\cite{Beers:1972xt}, the simplest way to build such a field dependent
algebra is to introduce the external field $A$ by a nonsingular transformation
\begin{equation}\label{defVV}
\Ad_{\VV(A)}:p_{1,3}\to p_{1,3}^d(A)=\VV(A)p_{1,3}\VV^{-1}(A).
\end{equation}
In case of a particular external electromagnetic field $A$, the external field
can be introduced by using an evolution operator $\VV(A)$, called the
``dynamical'' representation~\cite{Saar:1999ez,Ots:2001xn}. By analogy with the
free particle case one can realize this representation on the solution space of
relativistically invariant equations. Expressing the operators explicitly in
terms of free-field operators, one obtains the ``dynamical'' interaction.
Applying for instance the operator $\VV(A)$ to Eq.~(\ref{eq10}) one obtains
\begin{equation}\label{eq3.1}
\VV(A):\DD(\partial)\psi(x)=0\quad\rightarrow\quad\DD^d(\partial,A)\Psi(x,A)=0
\end{equation}
where $\DD^d(\partial,A)=\VV(A)\DD(\partial)\VV^{-1}(A)$ and
\begin{equation}\label{eq3.2}
\Psi(x,A)=\VV(A)\psi(x)
\end{equation}
(here and in the following we will skip the argument $x$ for $\Psi$ and the
argument $\partial$ for $\DD^d$). Having introduced the external gauge field
$A$, we introduce gauge covariance on the same foundation as Lorentz covariance
in Eq.~(\ref{eq01}), i.e.\ by claiming that the diagram
\begin{center}\begin{tabular}{rlcr}
$\Psi:$&$A$&$\longrightarrow$&$\Psi(A)$\\
$\llap{$g(\lambda)$}\downarrow\ $&$\ \downarrow\rlap{$\lambda$}$
  &&$\downarrow\rlap{$G(\lambda)$}$\\
$\Psi^\lambda:$&$A^\lambda=A+\partial\lambda$&$\longrightarrow$
  &$G(\lambda)\Psi(A)$\\
\end{tabular}\end{center}
is commutative, i.e.\
\begin{equation}\label{kom1}
\Psi^\lambda(A+\partial\lambda)=G(\lambda)\Psi(A).
\end{equation}
According to Eq.~(\ref{kom0}), the ``dynamical'' interaction $\DD^d$ is gauge
invariant under the gauge transformation
$A\to A^\lambda\equiv A+\partial\lambda$ if the diagram
\begin{center}\begin{tabular}{rlcr}
$\DD^d\Psi:$&$A$&$\longrightarrow$&$\DD^d(A)\Psi(A)$\\
$\downarrow\quad$&
$\ \downarrow\rlap{$\lambda$}$&&$\downarrow\rlap{$G(\lambda)$}\qquad$\\
$\DD^d\Psi^\lambda:$&$A+\partial\lambda$&$\longrightarrow$
  &$G(\lambda)\DD^d(A)\Psi(A)$\\
\end{tabular}\end{center}
is commutative, i.e.\
\begin{equation}\label{kom2}
\DD^d(A+\partial\lambda)\Psi^\lambda(A+\partial\lambda)
  =G(\lambda)\DD^d(A)\Psi(A).
\end{equation}
Together with Eq.~(\ref{kom1}) we obtain
$\DD^d(A+\partial\lambda)G(\lambda)\Psi(A)=G(\lambda)\DD^d(A)\Psi(A)$ or
\begin{equation}\label{kom2p}
\DD^d(A+\partial\lambda)G(\lambda)=G(\lambda)\DD^d(A)
\end{equation}
on the $\psi$-space. Note that up to now we have not specified the explicit
shape of the finite dimensional representation $G:\lambda\to G(\lambda)$ of
the gauge group.

\section{Specifying $\VV(A)$ by gauge invariance}
At this point we specify $\VV(A)$ by two claims. Due to gauge symmetry as a
fundamental principle the dynamical transformation $\VV$ has to be compatible
with the gauge transformation. Therefore, we first claim the gauge invariance
in~Eq.~(\ref{kom2p}) not only for the operator $\DD^d$ but for the whole
dynamical Poincar\'e algebra $p_{1,3}^d(A)$,
\begin{equation}
p_{1,3}^d(A+\partial\lambda)G(\lambda)=G(\lambda)p_{1,3}^d(A).
\end{equation}
By using Eq.~(\ref{defVV}) and multiplying by $G(\lambda)^{-1}$ from the right
we obtain
\begin{equation}
\VV(A+\partial\lambda)p_{1,3}\VV^{-1}(A+\partial\lambda)
  =G(\lambda)\VV(A)p_{1,3}(G(\lambda)\VV(A))^{-1}.
\end{equation}
This means that the first claim is fulfilled if
\begin{equation}
\VV(A+\partial\lambda)=G(\lambda)\VV(A).
\end{equation}
On the other hand, with Eqs.~(\ref{eq3.2}) and~(\ref{kom1}) we obtain
\begin{equation}
\VV^\lambda(A+\partial\lambda)\psi(x)=G(\lambda)\VV(A)\psi(x)
\end{equation}
and, therefore, $\VV^\lambda=\VV$ on the $\psi$-space. To summarize, by the
first claim the gauge symmetry determines the gauge properties of $\VV(A)$
and, therefore, of the interacting field $\Psi(A)$.

\vspace{7pt}
The second claim is that the dynamical transformation operator $\VV(A)$ should
be of Lorentz type, i.e.\ for the generators $s_{\mu\nu}$ of the Poincar\'e
algebra $p_{1,3}$ one has 
\begin{equation}\label{eq2.2}
\VV(A)s^{\mu\nu}\VV^{-1}(A)={V^\mu}_\rho(A){V^\nu}_\sigma(A)s^{\rho\sigma}
\end{equation}
which is a local extension of Eq.~(\ref{adjact}). $V(A)=V(x,A)$ is the local
Lorentz transformation generated by the external field $A$ and obeying
\begin{equation}\label{eq2.3}
V_{\mu\rho}(A){V^\mu}_\sigma(A)
  =V_{\rho\mu}(A){V_\sigma}^\mu(A)=\eta_{\rho\sigma}.
\end{equation}
If such a local Lorentz transformation exists, the problem is solved.
Therefore, in the following we make the attempt to find explicit realizations
of the local Lorentz transformation $V_{\mu\nu}(A)$. It is hard to find the
Lorentz transformation $V_{\mu\nu}(A)$ in general. However, as first shown by
Taub~\cite{Taub:1969zz}, in the case of a plane-wave field we obtain
\begin{equation}\label{eq2.4}
V_{\mu\nu}(A)=\eta_{\mu\nu}-\frac{q}{k_P}G_{\mu\nu}
  -\frac{q^2}{2k_P^2}A^2k_\mu k_\nu
\end{equation}
where $q$ is the electric charge of the particle and
$G_{\mu\nu}=k_\mu A_\nu-k_\nu A_\mu$. The plane wave field $A_\mu=A_\mu(\xi)$,
$\xi=kx$ is characterized by its lightlike propagation vector $k_\mu$, $k^2=0$,
and its polarization vector $a^\mu$ such that $a^2=-1$ and $ka=0$. The operator
$k_P\equiv k_\mu P^\mu$ commutes with any other and has a special role in the
theory. For particles with nonzero mass one has $k_\mu P^\mu\ne 0$. Therefore,
for the plane wave the differential operator $1/k_P$ is local and well-defined
for the plane-wave solution $\psi_P$ of the Klein--Gordon equation. In all
other cases, $1/k_P$ is assumed to exist.

\vspace{7pt}
Note that the plane-wave solution of the Dirac equation was found more than 70
years ago by Volkov~\cite{Volkov:1939} and extended later on to a field of two
beams of electromagnetic radiation~\cite{SenGupta:1967,Pardy:2005kv}. However,
these approaches did not make use of the nonsingular transformation $\VV(A)$.
The realization of $\VV(A)$ can be achieved by the nonsingular transformation
$\VV(A)=\VV_0(A)\VV_s(A)$ where
\begin{eqnarray}\label{eq2.10}
\VV_0(A)&=&\exp\Bigg\{-i\int\frac{d\xi}{2k_P}(2q(AP)-q^2A^2)\Bigg\},\nonumber\\
\VV_s(A)&=&\exp\Bigg\{-\frac{iq}{2k_P}G_{\mu\nu}s^{\mu\nu}\Bigg\}.
\end{eqnarray}
It has to be mentioned that the evolution operator $\VV(A)$ may be chosen to
be $H$-unitary according to the representation $T$ in Eq.~(\ref{eq03}), i.e.\
\[\VV^\dagger(A)H=H\VV^{-1}(A).\]
Considering the nonsingular transformation of Dirac-type wave equation
\begin{equation}
\VV(A):(\beta^\mu P_\mu-m)\psi=0\quad\rightarrow\quad
(\Gamma^\mu(A)\Pi_\mu(A)-m)\Psi(A)=0,
\end{equation}
with the help of Eq.~(\ref{eq2.10}) the ``dynamical'' counterparts to the
operator $P_\mu=i\partial_\mu$ can be calculated to be
$\Pi_\mu(A)=\VV(A)P_\mu\VV^{-1}(A)$,
\begin{eqnarray}
P_\mu&\rightarrow&\Pi_\mu(A)=P_\mu+k_\mu\frac{q}{2k_P}(qA^2-2AP-\slF),\\[4pt]
P^2&\rightarrow&\Pi^2(A)=(P-qA)^2-q\slF\label{P2trans}
\end{eqnarray}
($\slF\equiv s^{\mu\nu}F_{\mu\nu}$) while the ``dynamical'' counterpart to
$\beta^\mu$ is given by $\Gamma^\mu(A)=\VV(A)\beta^\mu\VV^{-1}(A)$,
\begin{equation}
\Gamma^\mu(A)={V^\mu}_\nu(A)\beta^\nu=\beta_\mu-\frac{q}{k_P}
  \left(\frac{q}{2k_P}A^2k_\mu k_\nu+G_{\mu\nu}\right)\beta^\nu.
\end{equation}
In terms of $\Pi_\mu(A)$ and $\Gamma^\mu(A)$ we have
\begin{equation}
\DD^d(A)\Psi(A)=(\Gamma^\mu(A)\Pi_\mu(A)-m)\Psi(A)=0.
\end{equation}
However, expressed in terms of $D_\mu=P_\mu-qA_\mu$ and $\beta^\mu$, we obtain
\begin{equation}\label{eq3.5}
\DD^d(A)\Psi(A)\equiv\left(\beta^\mu D_\mu-\frac q{2k_P}\slk\slF-m
  \right)\Psi(A)=0
\end{equation}
where $\slk\equiv\beta^\mu k_\mu$. This interaction is non-minimal. However,
as we have shown before, it is determined completely by the claim of gauge
invariance.

\vspace{7pt}
Note that due to the antimutation of the $\gamma$-matrices, in the spin-1/2
case the dynamical interaction in Eq.~(\ref{eq3.5}) reduces to the minimal
coupling. However, in order to obtain the correct values of the gyromagnetic
factor, in some cases the (phenomenological) Pauli term
$\gamma_\mu\gamma_\nu F^{\mu\nu}$ has to be added by hand to the minimal
coupling of the Dirac equation (see also Ref.~\cite{Sakurai:1993}, p.~109). In
case of plane waves the exact solution of this (supplemented) Dirac equation
as given by Chakrabarti~\cite{Chakrabarti:1968zz} obeys the same gauge
invariance condition $\Psi(A+\partial\lambda)=G(\lambda)\Psi(A)$. This
property is found also in the book by Fried~\cite{Fried:1990}.

\vspace{7pt}
Finally, as a consequence of the explicit form~(\ref{eq2.10}), the associated
transformation of the evolution operator $\VV(A)$ under the local gauge
transformation for the plane wave field,
\begin{equation}\label{eq2.15}
A_\mu(\xi)\rightarrow A_\mu(\xi)+\partial_\mu\lambda(\xi)
\end{equation}
becomes
\begin{equation}\label{eq2.16}
\VV(A)\rightarrow\VV(A+\partial\lambda)
  =e^{-iq\lambda}\VV(A).
\end{equation}
As an example of higher spin, the spin-3/2 case is considered in detail in
Ref.~\cite{Saar:2009jq}. As it turns out, the Rarita--Schwinger spin-3/2
equation on the presence of a ``dynamical'' interaction is algebraically
consistent and causal.

\section{Conclusions}
As a consequence of gauge invariance and Lorentz type of $\VV(A)$ we obtain
\begin{enumerate}
\item the invariance of the wave function under gauge transformations,
\begin{equation}
\Psi^\lambda(A+\partial\lambda)=\VV^\lambda(A+\partial\lambda)\psi
  =\VV(A+\partial\lambda)\psi=\Psi(A+\partial\lambda)
\end{equation}
i.e.\ $\Psi^\lambda=\Psi$,
\item the explicit shape of $G(\lambda)$ in Eq.~(\ref{kom1}),
\begin{equation}
\Psi^\lambda(A+\partial\lambda)=\VV(A+\partial\lambda)\psi
  =e^{-iq\lambda}\VV(A)\psi=e^{-iq\lambda}\Psi(A),
\end{equation}
i.e.\ $G(\lambda)=e^{-iq\lambda}$,
\item the invariance of $\DD^d$ under gauge transformations from
Eq.~(\ref{kom2}) and
\begin{equation}
\DD^d(A+\partial\lambda)\Psi^\lambda(A+\partial\lambda)
  =\DD^d(A+\partial\lambda)e^{-iq\lambda}\Psi(A),
\end{equation}
i.e.\ $\DD^d(A+\partial\lambda)G(\lambda)=G(\lambda)\DD^d(A)$ on the
$\psi$-space,
\item the ``dynamical'' interaction for any spin as given by
\begin{equation}
\DD^d(A)\Psi(A)=\left(\beta^\mu D_\mu-\frac q{2k_P}\slk\slF-m\right)
  \Psi(A)=0
\end{equation}
being non-minimal but completely determined by gauge invariance,\\
thereby causing Poincar\'e symmetry,
\item as a consequence of Eq.~(\ref{P2trans}), the gyromagnetic factor in
the presence\\ of a ``dynamical'' interaction as being $g=2$ for any
spin~\cite{Ots:2001xn}.
\end{enumerate}
Let us close again with Hermann Weyl. In Ref.~\cite{Weyl:1919} he honestly
confessed: {\it``Die entscheidenden Folgerungen in dieser Hinsicht verschanzen
sich aber noch hinter einem Wall mathematischer Schwierigkeiten, den ich
bislang nicht zu durchbrechen vermag.''} ({\it``However, the crucial
consequences in this respect entrench oneself still behind a bank of
mathematical difficulties which up to now I am not able to penetrate.''})
We hope that our work breaks a small bay into this mathematical bank.

\subsection*{Acknowledgements}
The work is supported by the Estonian target financed project No.~0180056s09.
S.G.\ acknowledges support by the Deutsche Forschungsgemeinschaft (DFG)
under grant 436 EST 17/1/06.

\end{document}